\documentclass[12pt,preprint]{aastex}
\shorttitle{Bullet Cluster in cDE Models}
\shortauthors{Lee \& Baldi}
\begin{document}
\title{Can Coupled Dark Energy Speed Up the Bullet Cluster?}
\author{ Jounghun Lee\altaffilmark{1}, Marco Baldi\altaffilmark{2,3}}
\altaffiltext{1}{Astronomy Program, Department of Physics and Astronomy, FPRD, 
Seoul National University, Seoul 151-747, Korea; jounghun@astro.snu.ac.kr}
\altaffiltext{2}{Excellence Cluster Universe, Boltzmannstr.~2, D-85748 Garching, 
Germany ; marco.baldi@universe-cluster.de}
\altaffiltext{3}{University Observatory, Ludwig-Maximillians University Munich, 
Scheinerstr. 1, D-81679 Munich, Germany}
\begin{abstract}
It has been recently shown that the observed morphological properties of the 
Bullet Cluster can be accurately reproduced in hydrodynamical simulations only 
when the infall pairwise velocity $V_{c}$ of the system exceeds $3000$ km/s (or 
at least possibly $2500$ km/s) at the pair separation of $2R_{\rm vir}$, where 
$R_{\rm vir}$ is the virial radius of the main cluster, and that the probability 
of finding such a bullet-like system is extremely low in the standard 
$\Lambda$CDM cosmology. We suggest here the fifth-force mediated by a coupled 
Dark Energy (cDE) as a possible velocity-enhancing mechanism and 
investigate its effect on the infall velocities of the bullet-like systems 
from the {\small CoDECS} (COupled Dark Energy Cosmological Simulations) public 
database. Five different cDE models are considered: three with constant coupling 
and exponential potential, one with exponential coupling and exponential 
potential, and one with constant coupling and supergravity potential.  For each 
model, after identifying the bullet-like systems, we determine the probability 
density distribution of their infall velocities at the pair separations of 
$(2-3)R_{\rm vir}$. Approximating each probability density distribution as a 
Gaussian, we calculate the cumulative probability of finding a bullet-like system 
with $V_{c}\ge 3000$ km/s or $V_{c}\ge 2500$ km/s. Our results show that in all 
of the five cDE models the cumulative probabilities increase compared to the 
$\Lambda$CDM case and that in the model with exponential coupling 
$P(V_{c}\ge \, 2500\, {\rm km/s})$ exceeds $10^{-4}$. 
The physical interpretations and cosmological implications of our results are 
provided. 
\end{abstract}
\section{INTRODUCTION}

The splendid success of the standard $\Lambda$CDM ($\Lambda$-Cold Dark Matter) 
cosmology that has been witnessed for the past two decades seems to be 
overshadowed by the recent discoveries of several possible anomalies 
\citep[e.g., see][for a review]{puzzle}. The so-called ``Bullet 
Cluster" (1E0657-56) is one of those observational challenges that  
the standard $\Lambda$CDM cosmology has to face \citep{MB08,LK10,TN11,AY11}. 
When \citet{tucker-etal95} first observed it while looking for a failed cluster, 
it appeared as a large cloud of hot gas. Later, the {\it Chandra} observation 
revealed that it is in fact a very rare system  composed of two head-on colliding 
massive clusters at $z=0.296$ in which a bullet-like subcluster is in the middle 
of escaping its main cluster's potential well at a bow-shock speed of 
$\sim 4700$ km/s after the first core passage
\citep{markevitch-etal02,markevitch-etal04,clowe-etal06,markevitch06}.

According to the results of \citet{MB08} -- based on high-resolution 
hydrodynamical simulations -- the pairwise infall velocity $V_{c}$ of a ``bullet" 
satellite onto the main cluster at a separation of $2R_{\rm vir}$ 
(where $R_{\rm vir}$ is the virial radius of the main halo) has to 
exceed $3000$ km/s, or possibly at least $2500$ km/s, in order to reproduce the 
peculiar morphological properties of the Bullet Cluster, such as the mass ratio 
between the two colliding clusters, the large separation between the gas and 
CDM distributions, and the high shock speed inferred from X-ray temperature 
measurements \citep[see also][]{SF07}. In the light of this result, \citet{LK10} 
investigated how probable it is for a ``bullet-like" system to have an infall 
velocity as high as $3000$ km/s, using the cluster catalogs from the large-volume 
MICE simulations \citep{mice10}. They found that the probability of finding a 
bullet-like system with $V_{c}\ge 3000$ km/s in a WMAP7 universe \citep{wmap7} is 
between $\sim 10^{-9}$ and $\sim 10^{-11}$, which led them to conclude that the 
existence of the Bullet Cluster is incompatible with the $\Lambda$CDM cosmology. 
Very recently, \citet{TN11} and \citet{AY11} have confirmed the results of 
\citet{LK10} and of \citet{MB08} using the data from different N-body and 
hydrodynamical simulations, respectively.

Both of a conservative and a radical approach to the Bullet Cluster problem have 
been recently proposed. The conservative one sought for a lower infall velocity 
solution in a $\Lambda$CDM cosmology under the suspicion that the results from 
the hydrodynamical simulations or the high shock velocity inferred 
from the X-ray temperatures may not be trustworthy \citep[e.g.,][]{fr-etal10}. 
In contrast, the radical approach attempted to figure out a mechanism capable of 
enhancing the infall velocities of the bullet-like systems in non-standard 
cosmologies. For example, \citet{WK10} claimed that in models with cascading 
gravity the probability of a bullet-like system with $V_{c}\ge 3000$ km is four 
orders of magnitude higher than in the $\Lambda$CDM model 
\citep[see also][]{MT10}.  Yet, their result was based on a pure analytical 
speculation without any numerical backup. 

Taking the radical direction, we examine here the possibility that the 
morphological properties and the required high infall velocity of the Bullet 
Cluster can be attributed to the presence of coupled Dark Energy (cDE). In cDE 
models, the Dark Energy is a scalar field $\phi$ with potential $U(\phi)$, which 
interacts with the CDM particles obeying the following two equations 
\citep{wetterich95,amendola00,amendola04}:
\begin{equation}
\label{eqn:cDE}
\ddot{\phi } + 3H\dot{\phi } +\frac{dU}{d\phi} = \sqrt{\frac{2}{3}}\beta (\phi ) 
\frac{\rho _{c}}{M_{{\rm Pl}}}\ , \qquad
\dot{\rho }_{c} + 3H\rho _{c} = -\sqrt{\frac{2}{3}}\beta (\phi )
\frac{\rho _{c}\dot{\phi }}{M_{{\rm Pl}}} \,,
\end{equation}
where an overdot represents a derivative with respect to the cosmic time $t$, 
$H\equiv \dot{a}/a$ is the Hubble function, $M_{\rm Pl}\equiv 1/\sqrt{8\pi G}$ is 
the reduced Planck mass, and $\rho _{\rm CDM}$ is the CDM density.
The interaction between the DE scalar field $\phi$ and the CDM fluid determines 
an exchange of energy-momentum, and a consequent time variation of the CDM 
particle mass $m_{\rm CDM}$. In Equations~(\ref{eqn:cDE}) the strength of the 
coupling  is fully determined by the coupling function $\beta (\phi)$, defined as 
$\beta (\phi )\equiv -d\ln m_{\rm CDM}/d\phi$. 

An enhancement of the infall velocity of the Bullet Cluster is naturally expected 
in cDE models since a long-range fifth-force generated by the coupling between 
$\phi$ and CDM particles has been found to play the role of accelerating 
structure formation processes 
\citep[e.g.,][and references therein]{mangano-etal03,maccio-etal04,MB06,PB08,
baldi-etal10,WP10}. The level of the enhancement of the infall velocity of the 
Bullet Cluster will however significantly depend on the shape of the scalar 
self-interaction potential $U(\phi)$ as well as the coupling 
function $\beta (\phi)$. We refer to the wide literature on the subject 
\citep[see e.g.,][and references therein]{amendola00,amendola04,PB08,
baldi-etal10,BLM11,baldi11a} for a more thorough description of the main 
basic features of cDE cosmologies. In particular, the specific models discussed 
in the present paper have been fully defined and presented by \citet{baldi11c} 
and \citet{codecs}.

The goal of this Paper is to study comprehensively the effect of cDE models on 
the infall velocities of the bullet-like systems analyzing the public cluster 
catalogs of the {\small CoDECS} project \citep{codecs} -- the largest suite of 
cosmological N-body simulations for cDE models to date --
which includes various choices for the scalar self interaction potential 
$U(\phi)$ and the coupling function $\beta (\phi)$.
The Paper is organized as follows. In \S 2, we provide a brief overview of the 
N-body simulations for the cDE models and describe how to select the bullet-like 
systems from the halo catalogs. In \S 3, we determine the probability density 
distribution of the bullet-like systems and calculate the probability of finding 
a bullet cluster with infall velocity larger than $3000$ km/s 
(and $2500$ km/s as well) for each cDE model. In \S 4, we provide the physical 
interpretation of our results and discuss its cosmological implications. 

\section{DATA AND ANALYSIS}

\subsection{A Brief Overview of the CoDECS}

The {\small CoDECS} (Coupled Dark Energy Cosmological Simulations) project is the 
largest suite of N-body simulations for cDE models ever performed, and all its 
numerical outputs, including halo and sub-halo catalogs, have been recently
made publicly available \citep{codecs}. Using the specific modified version by 
\citet{baldi-etal10} of the widely used parallel Tree-PM N-body code 
{\small GADGET} \citep{gadget2}, the different runs of the {\small L-CoDECS} 
suite follow the evolution of $1024^{3}$ CDM and $1024^{3}$ 
baryonic particles in a periodic cosmological box of linear comoving size 
$1\,h^{-1}$Gpc from $z_{i}=99$ to $z=0$ for different realizations of the cDE 
scenario. The individual CDM and baryonic particles have a mass of 
$5.84\times 10^{10}\,h^{-1}M_{\odot}$ and $1.17\times 10^{10}\,h^{-1}M_{\odot}$, 
respectively, at $z=0$, and the gravitational softening was set at a twenty-fifth 
of the mean linear interparticle spacing, $\epsilon_{g} = 20\,h^{-1}$kpc. 

The {\small CoDECS} suite includes -- besides the standard ``fiducial" 
$\Lambda $CDM scenario -- five different cDE cosmologies: four models with an 
exponential self-interaction potential \citep{LM85,RP88,wetterich88}
\begin{equation}
\label{eqn:exponential}
U(\phi ) = Ae^{-\alpha \phi }\,,
\end{equation}
and one model with a SUGRA potential \citep{BM99}
\begin{equation}
\label{eqn:SUGRA}
U(\phi ) = A\phi ^{-\alpha }e^{\phi ^{2}/2} \,,
\end{equation}
where for simplicity the scalar field has been redefined in units of the reduced 
Planck mass $M_{\rm Pl}$. All the former models have the same value for the 
potential slope $\alpha = 0.08$ but differ from one another in the 
coupling function: three models (named EXP001, EXP002, and EXP003) are 
characterized by a constant coupling $\beta (\phi) = {\rm const.}$ with values 
$0.05\,, 0.1\,,$ and $0.15$, respectively, consistent with present 
observational bounds on the DE-CDM interaction 
\citep[see e.g.,][]{bean-etal08,xia09,BV10}, while the remaining one (named 
EXP008e3) has an exponential coupling of the form $\beta (\phi)=0.4\cdot 
e^{3\phi}$.
For the latter, instead, a negative constant coupling $\beta = -0.15$ is assumed, 
allowing for a non-standard evolution of the resulting cDE model as proposed and 
explained in detail by \citet{baldi11c}. The combination of a SUGRA potential 
and of a negative coupling determines a peculiar dynamics of the scalar field,
that changes its direction of motion at some intermediate redshift between 
$z_{\rm CMB} \approx 1100$ and the present time. As a consequence of this 
inversion of motion, which for the specific model presented here (named SUGRA003) 
occurs at $z_{\rm inv} \approx 6.8$, the DE equation of state parameter 
$w_{\phi }$ shows a ``bounce" on the cosmological constant barrier 
$w_{\phi }=-1$. For this reason this new class of cDE models has been called the 
``Bouncing cDE scenario" \citep{baldi11c}.

All the models included in the {\small CoDECS} suite have been normalized 
according to the latest results from WMAP7 \citep{wmap7} both for what concerns 
their background cosmological parameters at $z=0$ and the amplitude of linear 
density perturbations at the last scattering surface $z_{\rm CMB}\approx 1100$. 
The {\small CoDECS} runs are therefore fully consistent with present bounds on 
the perturbations amplitude at $z_{\rm CMB}$ and allow to investigate the effects 
that a specific cDE scenario imprints on structure formation processes from 
$z_{\rm CMB}$ to the present. Table \ref{tab:model} lists the six cosmological 
models included in the {\small CoDECS} suite, with the corresponding coupling 
function, potential type, and the expected value of $\sigma _{8}$ that are 
derived for each scenario with the help of the linear perturbation theory. For a 
more detailed introduction to the {\small CoDECS} project and its related public 
database, see \citet{codecs}.

It is worth mentioning here that all the cDE models considered here, with the 
only exception of the SUGRA003 model, could be directly constrained by local 
measurements of the linear perturbations amplitude $\sigma _{8}$ at low 
redshifts. In particular, the EXP003 model might be already in tension with
presently available constraints on the power spectrum amplitude, 
$0.78\le\sigma_{8}\le 0.86$, which have been determined by several independent 
low-$z$ observations such as galaxy-galaxy correlation function, cosmic shear 
statistics, X-ray cluster abundance and Ly-$\alpha$ forest 
\citep[e.g.,][and references therein]
{mcdonald-etal05,hetter-etal07,henry-etal09,wen-etal10}. 
On the other hand, the EXP002 and EXP008e3 models appear still marginally 
consistent with this current limit on $\sigma_{8}$.

\subsection{Selecting the Bullet-Like Systems}

Using the {\small CoDECS} public halo catalogs at $z=0$, we first construct a 
mass-limited ($M\ge 10^{13}$ M$_{\odot}/h$) sample of the cluster-sized halos 
identified in the simulations of each cosmological model. The {\small CoDECS} 
halo catalogs have been produced by means of a Friends-of-Friends (FoF) 
algorithm \citep{FoF02} with linking length $\lambda = b\times \bar {d}$,
where $\bar{d}$ is the mean inter-particle separation and the linking parameter 
$b=0.2$ \citep{codecs}. Reapplying the FoF algorithm with a linking parameter of 
$b=0.33$ to the mass-limited sample from the CoDECS halo catalogs, we find the 
clusters-of-clusters for each of which the most massive cluster is identified as 
the main cluster while the others are considered as satellites, as done in 
\citet{LK10}.

The bullet-like systems are selected from the identified clusters of clusters 
according to the same four criteria that were used by \citet{LK10}:  $i)$ the 
main cluster's mass $M_{h}$ exceeds a certain cut-off value; $ii)$ the main 
cluster and at least one of its satellites are on the way of head-on collision 
($\vert\cos\alpha\vert\le 0.9$, where $\alpha$ is the angle between the 
velocities of the main cluster and the colliding satellite); $iii)$ the mass 
ratio between the satellite and main cluster is smaller than one fifth 
($M_{s}/M_{h}\le 1/5$); $iv)$ the satellite is located within the distances of 
$(2-3)R_{\rm vir}$ from its main cluster where $R_{\rm vir}$ is the virial radius 
of its main cluster. The bullet-like systems must satisfy the first three 
criteria to match the observed properties of the Bullet Cluster (IE0657-57) while 
the last criterion is used to find the infall pairwise velocities of the 
main-satellite cluster pairs before the collisions. For the detailed explanation 
of these criteria, we refer the readers to \citet{LK10}. 

Regarding the cut-off value of the main cluster's mass, it is worth mentioning 
here that the main cluster in the observed bullet cluster (IE0657-56) is a very 
massive one with mass of  $M_{h}\ge 10^{15}\,h^{-1}M_{\odot}$.  
In the current {\small CoDECS} samples, however, there are too few (less than 
$50$) bullet-like systems satisfying $M_{h}\ge 10^{15}\,h^{-1}M_{\odot}$. To 
avoid the poor number statistics, we use lower cut-off values, 
$M_{h}\ge 0.5\times 10^{15}\,h^{-1}M_{\odot}$ and 
$M_{h}\ge 0.7\times 10^{15}\,h^{-1}M_{\odot}$. We also focus on the $z=0$ sample 
for the same reason (at higher redshifts there are too few bullet-like systems). 

Table \ref{tab:nbullet} lists the number of clusters 
($N_{c}$) in the mass-limited sample, the number of the identified clusters of 
clusters ($N_{sc}$), the number and mean pairwise infall velocity of the selected 
bullet-like systems ($N_{bullet}$ and $\bar{V}_{c}$) for the two different cases 
of the main cluster's mass cutoff. As can be seen, the cDE models have 
systematically larger values of $N_{c},\ N_{sc},\ N_{bullet}$ than the 
$\Lambda$CDM model, which is consistent with the picture that the dark sector 
coupling leads to a faster growth of the large-scale structures. Regarding the 
mean infall velocities, all of the cDE models except of the SUGRA003 model 
yield higher values of $\bar{V}_{c}$. The larger the coupling is, the higher is 
the mean infall velocity of the bullet-like systems. In the SUGRA003 model, 
however, the value of $\bar{V}_{c}$ is slightly lower than the $\Lambda$CDM case 
when $M_{h}\ge 0.5\times 10^{15}\,h^{-1}M_{\odot}$, which should be attributed to 
the negative value of the coupling constant (see \S 4 for more discussions).

\section{INFALL VELOCITIES OF THE BULLET-LIKE SYSTEMS IN cDE MODELS}

For each selected bullet-like system, we measure the relative pairwise 
velocities, ${\bf V}_{c}\equiv {\bf V}_{h}-{\bf V}_{s}$, where ${\bf V}_{h}$ 
and ${\bf V}_{c}$ denote the velocities of the main and satellite clusters, 
respectively. Then, we bin the values of $\log V_{c}$ and count the number of 
those bullet-like systems with $\log V_{c}$ belonging to a given differential 
bin, $[\log V_{c},\ \log V_{c}+d\log V_{c}]$, to derive the probability density 
distribution of $\log V_{c}$ for each cDE model and for the $\Lambda$CDM model as 
well, as done in \citep{LK10}.
 
\begin{figure}[ht]
\begin{center}
\plotone{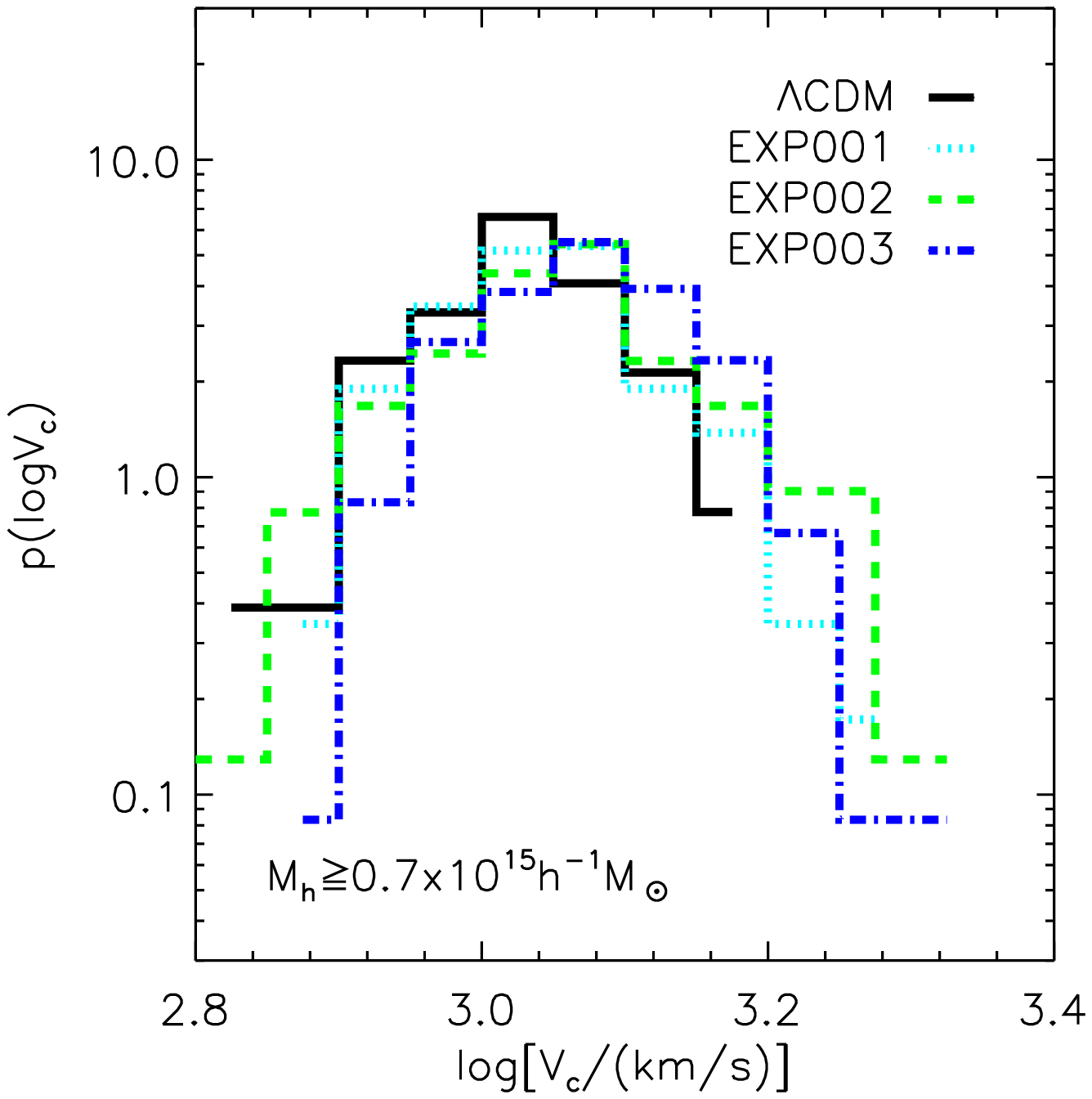}
\caption{Probability density distribution of the infall velocities, $\log V_c$, 
of the bullet-cluster-like  systems measured within $2\le r/R_{200}\le 3$ at 
$z=0$ for the four different dark energy models: $\Lambda$CDM, EXP001, EXP002, 
and EXP003 as dashed, dotted, solid and dot-dashed histograms, respectively. 
The main cluster masses are $M_{\rm main}\ge 0.7\times 10^{15}~h^{-1}~M_\sun$, 
for each case.}
\label{fig:pro1}
\end{center}
\end{figure}

Figure \ref{fig:pro1} plots $p(\log V_{c})$ at $z=0$ for the case of the standard 
$\Lambda$CDM cosmology (solid black histogram) and for the cases of the three cDE 
models with constant coupling and exponential potential (EXP001, EXP002 and 
EXP003 as dotted cyan, dashed green and dot-dashed blue histograms, 
respectively). As can be seen, the distribution $p(\log V_{c})$ 
shows a tendency to develop a longer high-velocity tail in the cDE models  with 
respect to $\Lambda$CDM. More specifically, the EXP002 model with $\beta=0.1$ 
exhibits the longest high-velocity tail among the three models. 

\begin{figure}
\begin{center}
\plotone{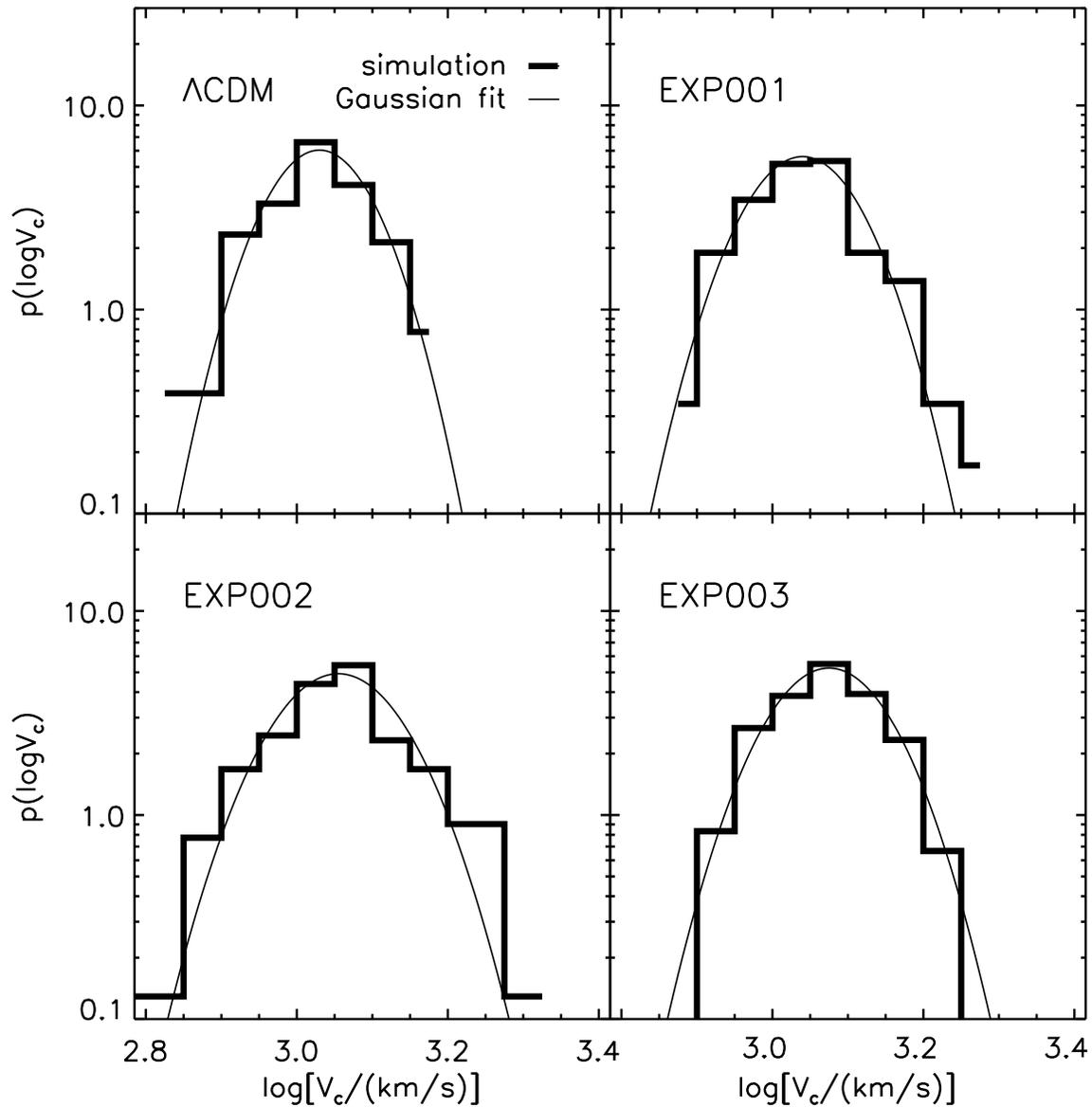}
\caption{Gaussian-fit (thin solid line) to the probability density distribution 
of the infall velocities, $\log V_c$, of the bullet-cluster-like  systems 
measured within $2\le r/R_{200}\le 3$ at $z=0$ (thick solid histogram) for the 
four different dark energy 
models: $\Lambda$CDM, EXP001, EXP002, and EXP003 in the top-left, top-right, 
bottom-left and bottom-right panel, respectively.  The main cluster masses are 
$M_{\rm main}\ge 0.7\times 10^{15}~h^{-1}~M_\sun$, for each case.}
\label{fig:fit1}
\end{center}
\end{figure}

To quantify the difference among the four distributions and to calculate the 
cumulative probability of $V_{c}$, we fit $p(\log V_{c})$ to a Gaussian 
distribution by adjusting for each model the mean and standard deviation, 
$(\nu,\ \sigma_{\nu})$, as done in \citet{LK10}. Figure \ref{fig:fit1} plots the 
fitting results. As can be seen, the Gaussian distribution gives a reasonably 
good fit to $p(\log V_{c})$ for each case. Table \ref{tab:fit} lists the best-fit 
values of $(\nu,\ \sigma)$ for the three models, showing that the Gaussian fit to 
the infall velocity distribution $p(\log V_{c})$ has a larger mean and a larger 
standard deviation in the constant coupling cDE models than in the $\Lambda$CDM 
cosmology. The infall velocity distribution with the highest mean is found, as 
expected, for the case of the EXP003 model, while EXP002 has the largest standard 
deviation.

\begin{figure}
\begin{center}
\plotone{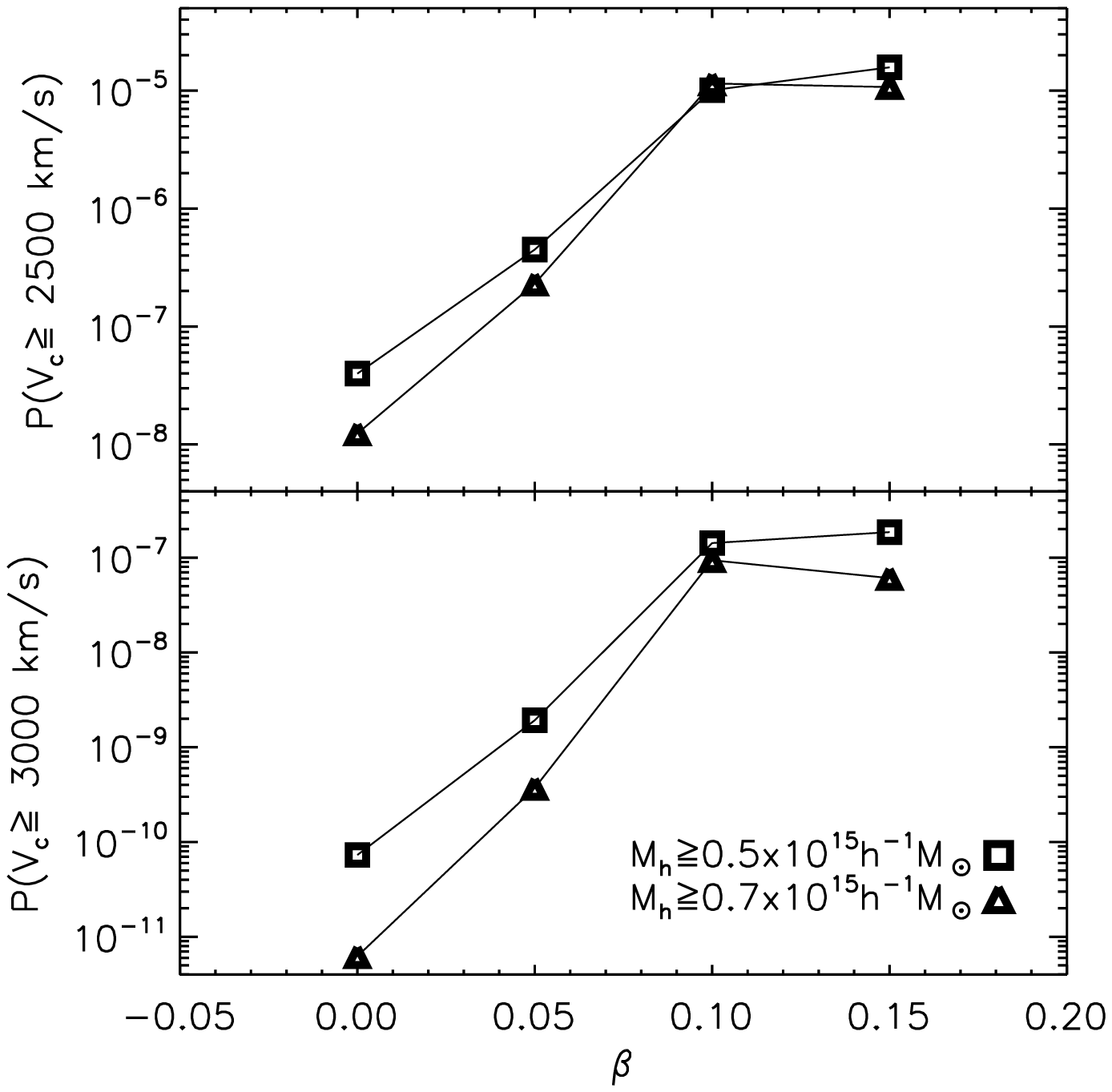}
\caption{Cumulative probabilities of the infall velocities of the bullet clusters 
versus the cDE coupling constant, $\beta$ for the $\Lambda$CDM, EXP001, EXPOO2, 
and EXP003 models.}
\label{fig:cpro1}
\end{center}
\end{figure}

Integrating the best-fit Gaussian distributions, we also compute for each model 
the cumulative probabilities, $P(V_{c}\ge 3000\, {\rm km/s})$ and 
$P(V_{c}\ge 2500\, {\rm km/s})$, which are plotted in Figure \ref{fig:cpro1} as a 
function of the constant coupling $\beta$, where $\Lambda$CDM corresponds to 
$\beta=0$. As can be seen, the highest values of the two cumulative 
probabilities are found for the case of the EXP002 model. The values of 
$P(V_{c}\ge 3000\, {\rm km/s})$ and $P(V_{c}\ge 2500\, {\rm km/s})$ increase by a 
factor of $10^{3}$ and $10^{4}$ in the EXP002 model compared to $\Lambda$CDM 
case. It is interesting to note that although the EXP003 model has a stronger 
coupling, its cumulative probabilities are not higher than those of the EXP002 
model.

\begin{figure}[ht]
\begin{center}
\plotone{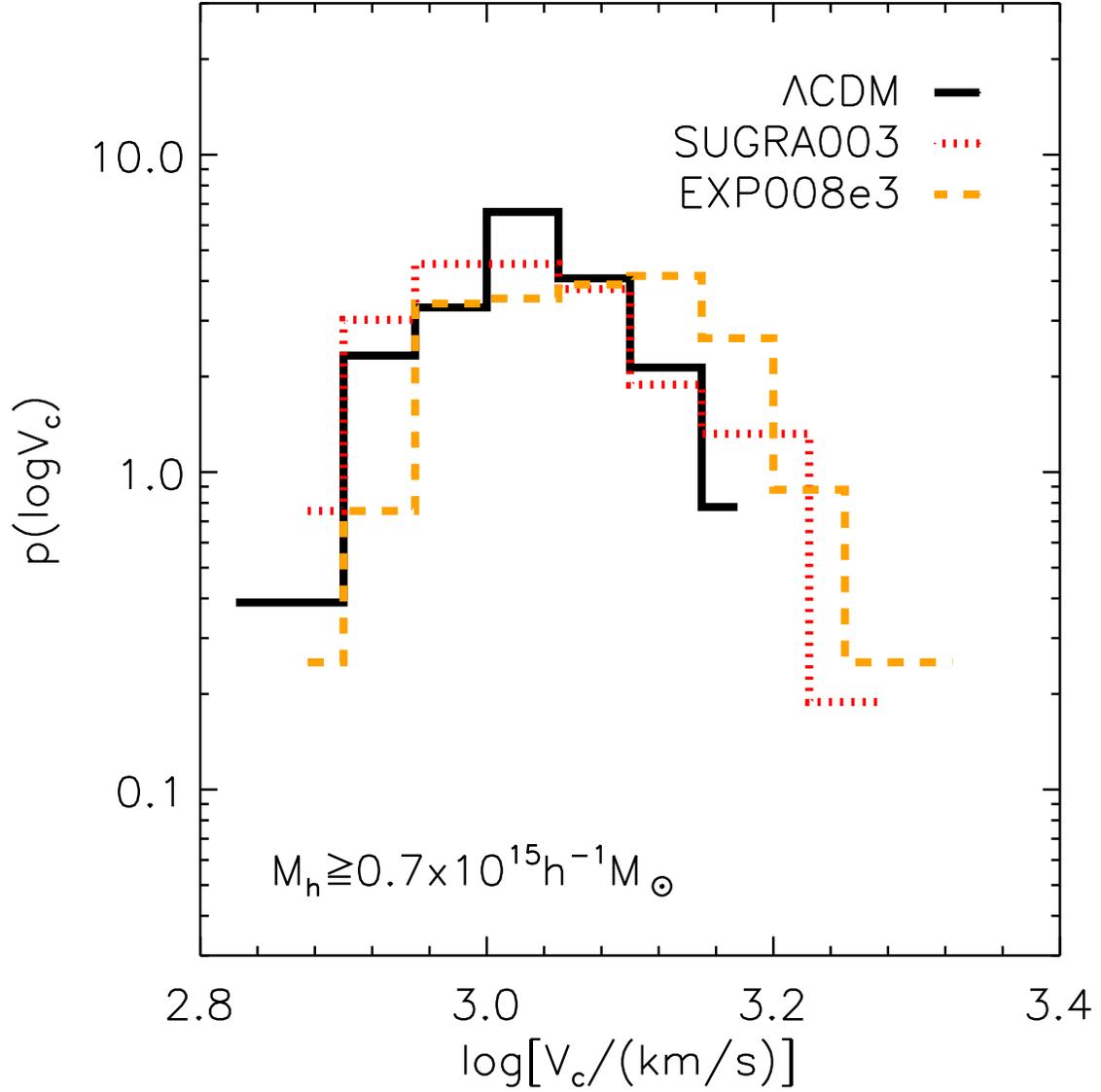}
\caption{Same as Figure \ref{fig:pro1} but for the SUGRA003 and EXP008e3 models 
(dotted and dashed histograms, respectively) in comparison with the $\Lambda$CDM 
case (solid histogram).}
\label{fig:pro2}
\end{center}
\end{figure}

\begin{figure}
\begin{center}
\plotone{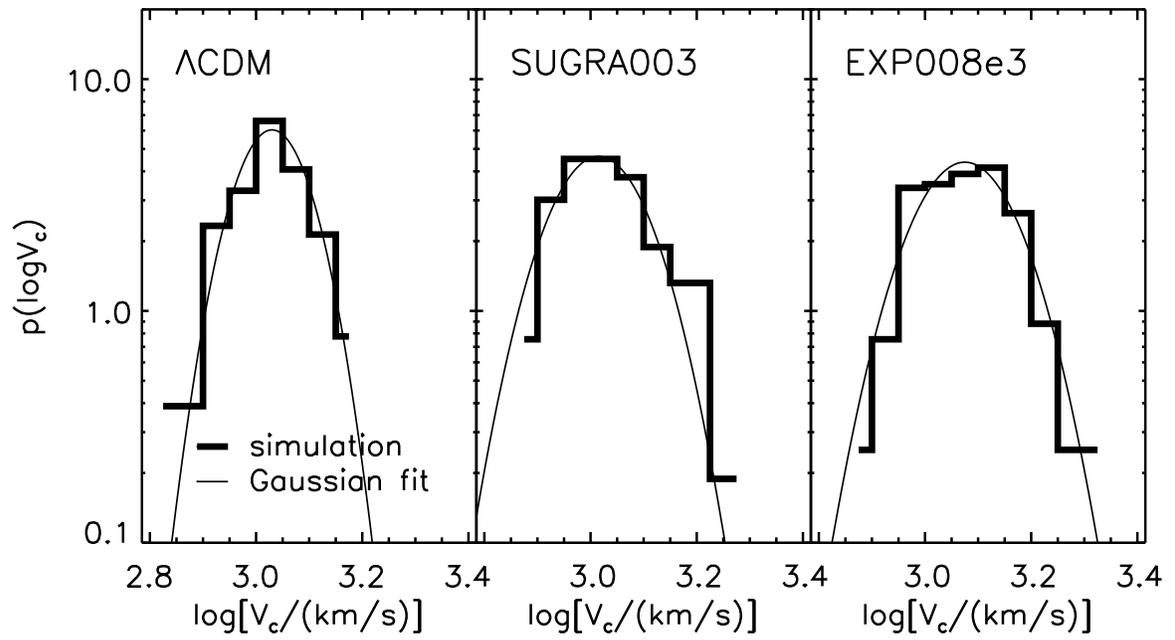}
\caption{Same as Figure \ref{fig:fit1} but for the SUGRA003 and EXP008e3 models 
in comparison with the $\Lambda$CDM case.}
\label{fig:fit2}
\end{center}
\end{figure}

The same type of analysis has been carried out for the other two cDE models 
included in the {\small CoDECS} suite, namely EXP008e3 and SUGRA003.
The results are shown in Figure \ref{fig:pro2}, where the probability density 
distribution $p(\log V_{c})$ computed for these models is plotted and 
and compared to the $\Lambda$CDM case, while Figure \ref{fig:fit2} plots the 
Gaussian fitting functions whose best-fit means and best-fit standard deviations 
are listed in Table \ref{tab:fit}. As can be seen, the probability density 
distribution $p(\log V_{c})$ for the case of the EXP008e3 model has 
both the largest mean and standard deviation, while the SUGRA003 model yields a 
smaller mean but a larger standard deviation of $p(\log V_{c})$ than 
$\Lambda$CDM. 

\begin{figure}
\begin{center}
\plotone{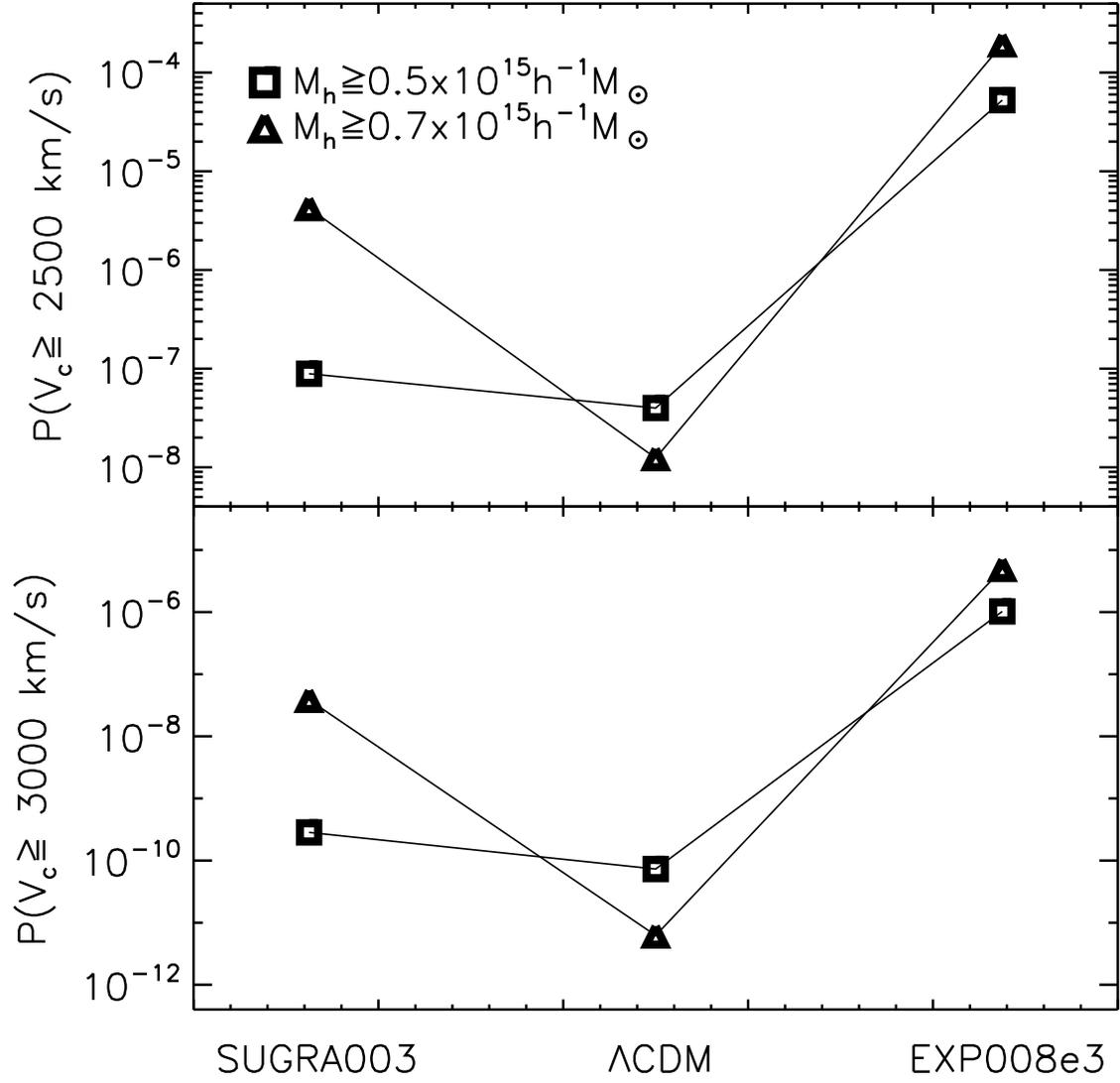}
\caption{Cumulative probabilities of the infall velocities of the bullet clusters 
versus the cDE coupling constant, 
$\beta$ for the SUGRA003 AND EXP008e3 models.}
\label{fig:cpro2}
\end{center}
\end{figure}

The cumulative probabilities, $P(V_{c}\ge 3000\, {\rm km/s})$ and 
$P(V_{c}\ge 2500\, {\rm km/s})$, are plotted in the top and bottom panels of 
Figure \ref{fig:cpro2}, respectively. As can be seen, both the SUGRA003 and 
EXP008e3 models have larger cumulative probabilities than $\Lambda $CDM. 
For the case of SUGRA003, when the lower mass limit for the main cluster is 
set at $0.7\times 10^{15}$ M$_{\odot}/h$, the cumulative probabilities, 
$P(V_{c}\ge 3000 {\rm km/s})$ and $P(V_{c}\ge 2500 {\rm km/s})$, increase 
by a factor of $10$ and $10^{2}$, respectively, compared to $\Lambda$CDM. 
However, when the mass cutoff is set at $0.5\times 10^{15}$ M$_{\odot}/h$, 
the SUGRA003 model shows almost no enhancement. In contrast, EXP008e3 exhibits 
the maximal enhancement, no matter which lower mass limit of the main cluster is 
used. The cumulative probabilities, $P(V_{c}\ge \ 3000\, {\rm km/s})$ and 
$P(V_{c}\ge \ 2500\, {\rm km/s})$, increase by a factor of $10^{5}$ and $10^{4}$, 
respectively, compared to the $\Lambda$CDM result. 
Note that in this model $P(V_{c}\ge \ 2500\, {\rm km/s})$ reaches up to 
above $10^{-4}$, which indicates that if the required infall velocity of a 
bullet-like system can be possibly reduced to $2500$ km/s \citep[][]{AY11}, then 
the existence of the Bullet Cluster is not such an extremely rare event in a cDE 
model like the exponential coupling scenario EXP008e3.

Furthermore, it is worth mentioning here that the above cumulative probabilities 
are likely to be underestimated due to the limited volume of the {\small CoDECS} 
runs. As shown in \citet{fr-etal10}, the probability of finding a bullet cluster 
increases almost linearly with the simulation box volume. In other words, the 
larger volume a simulation box has, the more probable it is to find a bullet-like 
system with required infall velocity. In fact, we find for the case of 
$\Lambda$CDM model $P(V_{c}\ge 3000\, {\rm km/s})\sim 10^{-12}$ which is an order 
of magnitude lower than what \citet{LK10} found, 
$P(V_{c}\ge 3000\, {\rm km/s})\sim 10^{-11}$, from the MICE simulation of volume 
$27\,h^{-3}{\rm Gpc}^{3}$. Given this, we suspect that the real value of 
$P(V_{c}\ge 3000\, {\rm km/s})$ and $P(V_{c}\ge 2500\, {\rm km/s}$) would be 
higher than our estimates $10^{-6}$ and $10^{-4}$, respectively.

\section{DISCUSSION AND CONCLUSION}

We have carried out a systematic investigation of how the probability of finding 
a bullet-like system with the required high-infall velocity ($3000$ km/s or 
possibly at least $2500$ km/s) changes in the context of cDE models compared to 
the standard $\Lambda$CDM cosmology, by analyzing the public numerical data from 
{\small CoDECS}, the largest N-body simulations of cDE models to date, with a box 
size of $1$ comoving $h^{-3}$Gpc$^{3}$. Three models with constant coupling and 
exponential potential (EXP001, EXP002, EXP003), one bouncing cDE model with 
constant coupling and SUGRA potential (SUGRA003), and one model with exponential 
coupling and exponential potential (EXP008e3) are considered in the {\small 
CoDECS} suite, besides the fiducial $\Lambda$CDM cosmology. 

In all of the five cDE models the cumulative probabilities have been found to be 
higher than in $\Lambda$CDM. It has been also shown that the maximal four and 
five orders of magnitude enhancements of the cumulative probabilities are yielded 
by the EXP008e3 model where $P(V_{c}\ge 2500\, {\rm km/s})$ and 
$P(V_{c}\ge 3000\, {\rm km/s})$ reach up to above $10^{-4}$ and $10^{-6}$, 
respectively. This is a remarkable result, considering that all the 
{\small CoDECS} simulations share the same random phases in the initial 
conditions and the same amplitude of density perturbations at the last scattering 
surface. Furthermore, given the possible underestimate of the cumulative 
probabilities due to the limited volume of {\small CoDECS} \citep{fr-etal10}, we 
argue that our results should be considered conservative and that therefore 
the existence of the bullet cluster should not be regarded as an extremely rare 
event in the EXP008e3 model. Yet, in the other cDE models it should be still 
quite unlikely  to find a bullet cluster with the observed morphology.

The different amplitudes of the bullet velocity enhancement in the different cDE 
models can be qualitatively understood by considering the time evolution of the 
linear density and velocity perturbations of each scenario. By taking into 
account only the coupling strength and its integrated effect on the growth of 
density perturbations, in fact, our finding that the maximal velocity enhancement 
is obtained for the exponential coupling model (EXP008e3) might look surprising. 
As shown by \citet{codecs} (see the right panel of Figure 2 in the reference),  
the maximal effect on the density perturbation amplitude is obtained for the 
EXP003 model that reaches at $z=0$ an enhancement of $\delta(z)$ of about 20\% 
relative to $\Lambda $CDM, while the EXP008e3 model shows a $10\%$ enhancement 
(as also described by the different $\sigma _{8}$ values listed in Table 
\ref{tab:model} among the models). 

However, if we also consider the velocity perturbation variable $\theta$, 
defined as
\begin{equation}
\label{eqn:theta}
\theta (z)\equiv \frac{1}{\Omega _{\rm c} + \Omega _{b}}
\left( \theta _{c}\Omega _{c} + \theta _{b}\Omega _{b}\right)
\end{equation}
where $\theta _{b,c}\equiv \vec{\nabla}\cdot \vec{u}_{b,c}$ are the individual 
velocity perturbation variables, $\vec{u}_{b,c}$ are the peculiar velocities, 
and $\Omega _{b,c}$ the relative density parameters for baryons and 
CDM,respectively, we can separately solve for $\theta (z)$ in the different 
cosmologies, and compare the time evolution of the velocity perturbations to the 
$\Lambda $CDM case. 
In the linear regime, the density and velocity perturbations $\delta _{b,c}$
and $\theta _{b,c}$ for baryons and CDM in a cDE cosmology evolve according to 
the following system of coupled differential equations:
\begin{eqnarray}
\label{del_c}
\delta '_{c} &=& \theta _{c} - f(\beta _{c})\delta _{c}\\
\label{th_c}
\theta '_{c} &=& \left[ \frac{1}{2}\left( 3w_{\phi }\Omega _{\phi } + \Omega _{r} 
-1\right) + g(\beta _{c},\phi ')\right] \theta _{c} + \frac{3}{2}\left[ \Omega 
_{b}\delta _{b} + \Omega _{c}\delta _{c}\Gamma _{c}\right] \\
\label{delb}
\delta '_{b} &=& \theta _{b}\\
\label{th_b}
\theta '_{b} &=& \left[ \frac{1}{2}\left( 3w_{\phi }\Omega _{\phi } + \Omega _{r} 
-1\right)\right] \theta _{b} + \frac{3}{2}\left[ \Omega _{b}\delta _{b} + \Omega 
_{c}\delta _{c}\right]\,,
\end{eqnarray}
where a prime denotes a derivative with respect to the e-folding time $\alpha 
\equiv \ln a$, $\Omega _{\phi }$ and $\Omega _{r}$ are the DE and radiation 
fractional densities, respectively, $w_{\phi }$ is the DE equation of state 
parameter, and the three functions $f(\beta _{c})$, $g(\beta _{c},\phi ')$
and $\Gamma _{c}(\phi )\equiv 1 + 4\beta _{c}^{2}(\phi )/3$ encode the deviation 
from the standard $\Lambda $CDM cosmology due to the DE interaction, being 
$(f,g,\Gamma)=(0,0,1)$ for an uncoupled DE field \citep[see e.g.][for a more 
detailed description of the perturbation equations and a
definition of the functions $f$ and $g$]{amendola04}.

\begin{figure}
\begin{center}
\plotone{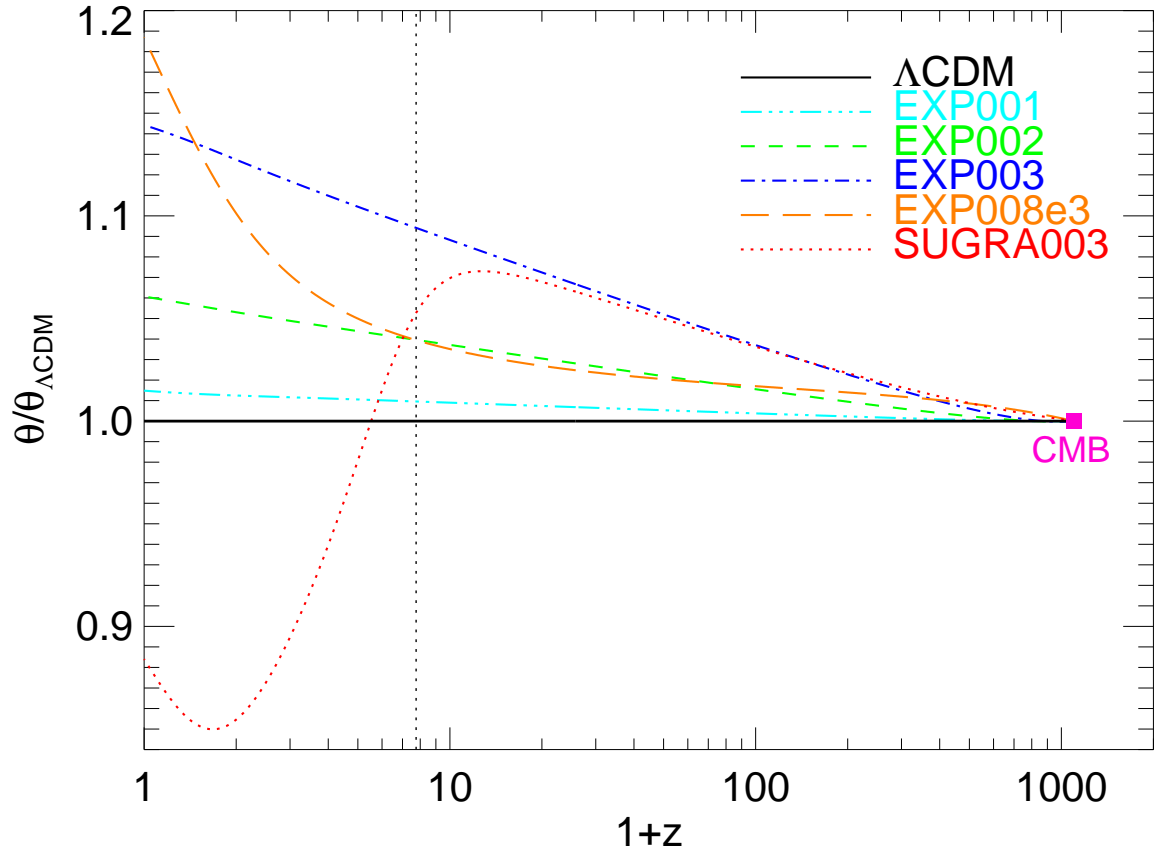}
\caption{Velocity perturbation $\theta$ for all the cDE models considered in the 
present work relative to the $\Lambda $CDM case $\theta _{\Lambda {\rm CDM}}$. 
The vertical dotted line indicates the redshift at which the Bouncing cDE 
model SUGRA003 features the inversion of the scalar field motion, corresponding 
to $z_{\rm inv}\approx 6.8$. Note that when this transition occurs at 
$z_{\rm inv}$ the trend of the linear velocity perturbation of the SUGRA003 model 
with respect to $\Lambda $CDM case is also inverted. 
On the other hand the fast growth of the coupling function at low redshifts for 
the EXP008e3 model leads to the steep increase in the velocity perturbation 
relative to the $\Lambda$CDM case. In consequence the EXP008e3 model 
has a larger value of $\theta/\theta_{\rm {\Lambda}CDM}$ at $z=0$ than 
the EXP003 model even though the overall amplitude of the linear density 
perturbation is higher in the EXP003 model.}
\label{fig:theta}
\end{center}
\end{figure}

By solving the system (\ref{del_c}-\ref{th_b}) we can separately compute the time 
evolution of the individual density and velocity perturbations for baryons and 
CDM and derive the total velocity perturbation $\theta $ given by 
Equation~(\ref{eqn:theta}).
In Figure~\ref{fig:theta} we plot the ratio of the velocity perturbation 
$\theta $ to the $\Lambda $CDM case for all the models considered in the present 
work. As one can see from the figure, the situation is significantly different 
with respect to the evolution of the density perturbation $\delta $.
In particular, we display here for the first time the redshift evolution of the 
velocity perturbation $\theta $ in a significant number of cDE scenarios, and
our results show how a given cDE model can have a significantly different 
relative impact on the density and velocity perturbations. Figure \ref{fig:theta} 
clearly shows that the maximal enhancement of velocity perturbations with respect 
to $\Lambda$CDM is realized by the EXP008e3 model (orange, long-dashed line) -- 
consistently with our findings on the infall velocity in the bullet-like systems 
identified in the {\small CoDECS} simulations -- even if the corresponding 
enhancement of the density perturbations amplitude (i.e. of $\sigma _{8}$) is 
significantly smaller than in the EXP003 model (blue, dot-dashed line).

Although the solution for $\theta$ displayed in Figure \ref{fig:theta}
is strictly valid only in the linear regime (as for the case of the density 
perturbation $\delta $), the relative impact of the different cDE models
on the velocity perturbations can be considered reliable also in the nonlinear 
regime for situations where the velocity vector is aligned with the gradient of 
the gravitational potential, since in such a case the extra-friction and the 
fifth-force contributions arising as a consequence of the DE interaction are also 
aligned with each other and keep behaving as for the linear case. 
This is in particular the situation of the bullet-like systems that we are 
considering in the present work, due to the head-on collision between the two 
clusters. Therefore we can consider the results obtained for the linear 
solution of the velocity perturbation $\theta $ as indicative of the impact of 
the coupling even in these highly nonlinear systems. See \citet{baldi11b} for a 
discussion on the linear and nonlinear effects of the friction term. 

Particularly interesting is also the case of the bouncing cDE model SUGRA003: for 
this scenario, in fact, the velocity perturbation $\theta $ follows a similar 
path as for the EXP003 model at high redshifts, but is suddenly slowed down in 
correspondence to the bounce of the DE scalar field at $z_{\rm inv}\approx 6.8$, 
and is the only model to have a smaller velocity perturbation $\theta$ as 
compared to $\Lambda $CDM at low redshifts. 
This is consistent with the background dynamics of the bouncing cDE scenario, 
that features an inversion of the scalar field motion at $z_{\rm inv}$ with the 
consequent change of sign of the friction term $\dot{\phi }\beta (\phi)$, that 
accelerates particles in their direction of motion before $z_{\rm inv}$, while it 
decelerates particles after $z_{\rm inv}$. The suppression of the velocity 
perturbation variable $\theta $ after $z_{\rm inv}$ is therefore an expected
effect due to the peculiar dynamics o the bouncing cDE scenario.
This is reflected in the lower value of the mean pairwise infall velocity 
detected in our sample for the Bouncing cDE model (see Table~\ref{tab:fit}), 
unique among the cDE models considered in our analysis.

Nevertheless, as already mentioned above, the effect of the friction term 
become more complex in the nonlinear regime, due to the relative
orientation between the local velocity field and the gravitational potential 
gradient. We speculate here that this additional degree of freedom that modulates 
the efficiency of the friction term in slowing down CDM particle is responsible 
for the larger variance detected in the pairwise infall velocity distribution in 
our sample (see again Table~\ref{tab:fit}) which is reflected in a slightly 
enhanced detection probability even for the SUGRA003 scenario, in spite of the 
reduction of the mean infall velocity.

\begin{figure}
\begin{center}
\plotone{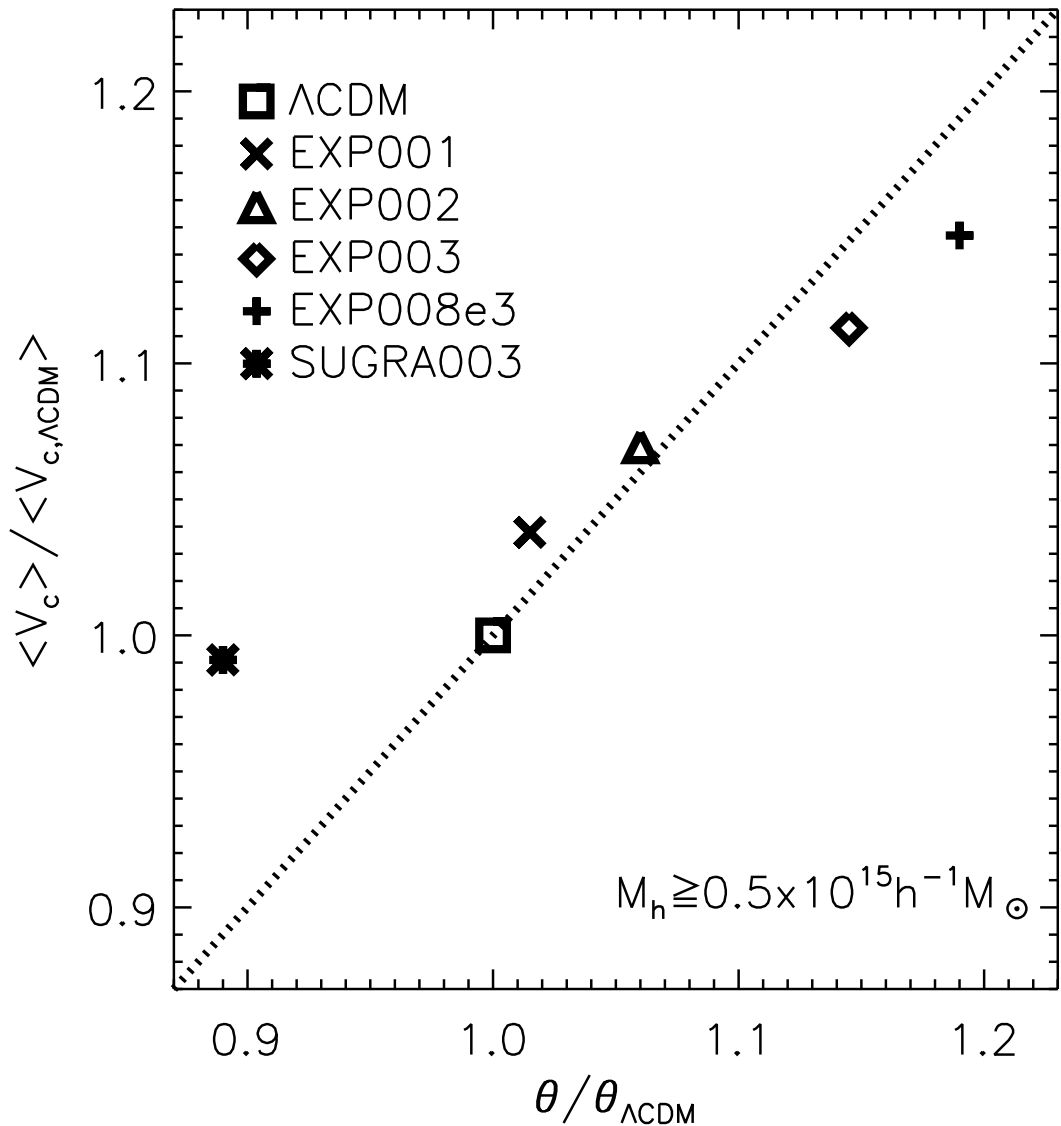}
\caption{Ratio of the mean infall velocities for all the models to that 
for the $\Lambda$CDM case versus the theoretical predicted ratio of the 
velocity perturbations for all models to that for the $\Lambda$CDM case. 
The mean infall velocities are obtained by taking the averages over the selected 
bullet-like systems with $M_{h}\ge 0.5\times 10^{15}\,h^{-1}M_{\odot}$. 
The dotted line corresponds to the case of 
$\langle V_{c}\rangle/\langle V_{\rm c,{\Lambda}CDM}\rangle =
\theta/\theta_{\rm \Lambda CDM}$.}
\label{fig:v05}
\end{center}
\end{figure}

\begin{figure}
\begin{center}
\plotone{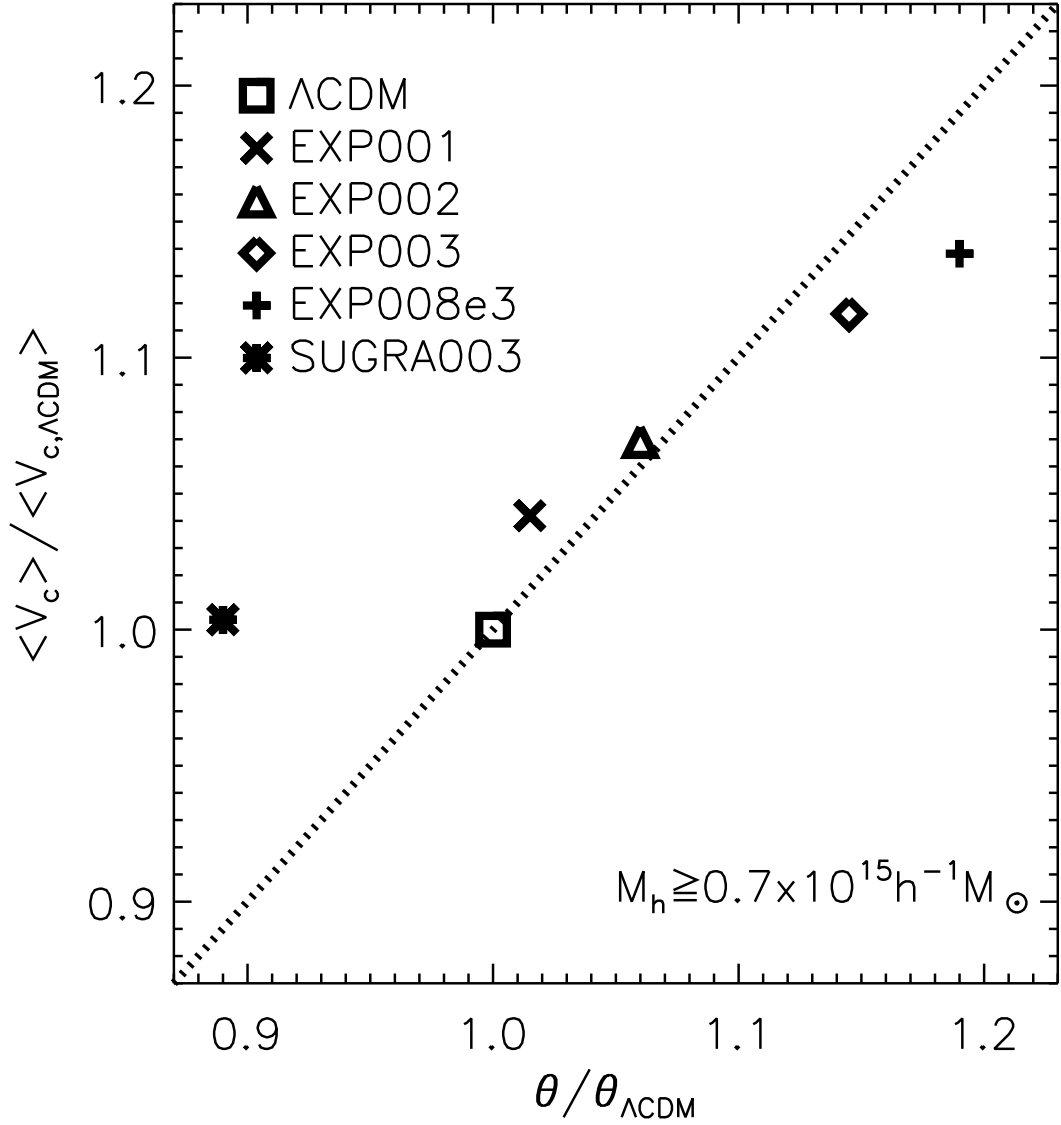}
\caption{Same as Figure \ref{fig:v05} but for the case of 
$M_{h}\ge 0.7\times 10^{15}\,h^{-1}M_{\odot}$.}
\label{fig:v07}
\end{center}
\end{figure}

To see whether or not the enhanced velocities of the bullet-like systems in 
cDE models are really related to the enhanced values  of the linear velocity 
perturbations,  we compare the ratio of the mean infall velocity of the bullet-
like systems for each cDE model to that for the $\Lambda$CDM model, 
$\bar{V}_{c}/\bar{V}_{\rm c,{\Lambda}CDM}$ (from Table \ref{tab:nbullet}) 
with the values of $\theta/\theta_{\rm {\Lambda}CDM}$ at $z=0$ 
(from Figure \ref{fig:theta}). The results are plotted in 
Figures \ref{fig:v05} and \ref{fig:v07} for the cases of 
$M_{h}\ge 0.5\times 10^{15}\,h^{-1}M_{\odot}$ and $M_{h}\ge 0.7\times 
10^{15}\,h^{-1}M_{\odot}$, respectively. In each Figure, the dotted line 
corresponds to the case that the two ratios have the same values. 
As can be seen, the enhancements of $\bar{V}_{c}$ relative to the 
$\Lambda$CDM case indeed follow the pattern of the enhancements of $\theta$ 
relative to $\theta_{\rm {\Lambda}CDM}$.  
In the SUGRA003 model, however, $\bar{V}_{c}/\bar{V}_{\rm c,{\Lambda}CDM}$ is  
significantly larger than $\theta/\theta_{\rm {\Lambda}CDM}$ at $z=0$, which 
reflects the peculiar evolution of the SUGRA003 velocity perturbations 
shown in Figure \ref{fig:theta}.

To conclude, we have performed a comprehensive investigation of the effect of 
coupled dark energy models on the pairwise infall velocity of the bullet-like 
systems identified in the public catalogs of the {\small CoDECS}
N-body simulations. Since our results have clearly demonstrated that coupled dark 
energy has the effect of speeding up the bullet-like systems and that the 
strength of the effect significantly depends on the shape of the scalar potential 
as well as of the coupling function,  we conclude that the Bullet Cluster will 
become a valuable constraint of the coupled dark energy scenario.

\acknowledgments

We thank an anonymous referee for useful comments.
J.L. acknowledges the financial support from the National Research Foundation 
of Korea (NRF) grant funded by the Korea government (MEST, No.2011-0007819) 
and from the National Research Foundation of Korea to the Center for Galaxy 
Evolution Research. M.B. is supported by the DFG Cluster of Excellence ``Origin 
and Structure of the Universe'' and by the TRR33 Transregio Collaborative 
Research Network on the ``Dark Universe''.

\clearpage
\clearpage
\clearpage
\clearpage
\clearpage
\clearpage

\clearpage
\clearpage
\clearpage
\clearpage
\begin{deluxetable}{ccccc}
\tablewidth{0pt}
\setlength{\tabcolsep}{5mm}
\tablecaption{Defining properties of the six cosmological models 
\citep[see][for the detailed descriptions of the models]{codecs}.}
\tablehead{Model & $U(\phi )$ & $\alpha $ & $\beta (\phi)$ & $\sigma_{8}$ } 
\startdata
$\Lambda $CDM  & ${\rm const.}$ & -- & -- & $0.809$ \\ 
EXP001  & $e^{-\alpha \phi }$ & $0.08$ & $0.05$ & $0.825$ \\ 
EXP002 & $e^{-\alpha \phi }$ &$0.08$ & $0.1$ & $0.875$\\  
EXP003  & $e^{-\alpha \phi }$ &$0.08$ & $0.15$ & $0.967$\\
EXP008e3 & $e^{-\alpha \phi }$ & $0.08$ & $0.4 e^{3\phi }$ & $0.895$ \\
SUGRA003 & $\phi ^{-\alpha }e^{\phi ^{2}/2}$ & $2.15$ & $-0.15$ & $0.806$
\enddata
\label{tab:model}
\end{deluxetable}
\clearpage
\begin{deluxetable}{ccccccc}
\tablewidth{0pt}
\setlength{\tabcolsep}{5mm}
\tablecaption{Model, number of clusters, number of clusters of clusters, number 
and mean infall velocity of bullet-like systems for the two different cases of the 
main cluster masses}
\tablehead{model & $N_{c}$ & $N_{sc}$ & $N_{\rm bullet}$\tablenotemark{a} 
& $N_{\rm bullet}$\tablenotemark{b} & $\bar{V}_{c}$\tablenotemark{a} 
& $\bar{V}_{c}$\tablenotemark{b} \\
& & & & & [km/s] & [km/s]} 
\startdata
$\Lambda$CDM  & $121936$ & $22019$ & $184$ & $103$ & $999.56$ & $1073.33$\\
EXP001  & $124467$ & $22448$ & $195$ & $116$ & $1037.38$ & $1118.27$\\ 
EXP002 & $131880$ & $23648$ & $231$ & $155$ & $1069.03$ & $1247.71$\\  
EXP003  & $141915$ & $25223$ & $357$ & $240$ & $1112.55$ & $1197.93$\\
EXP008e3 & $132992$ & $23795$ & $233$ & $159$ & $1146.49$ & $1221.66$\\
SUGRA003 & $127651$ & $23216$ & $188$ & $106$ & $990.53$ & $1077.21$\\  
\enddata
\tablenotetext{a}{For $M_{h}\ge 0.5\times 10^{15}\,h^{-1}M_{\odot}$.} 
\tablenotetext{b}{For $M_{h}\ge 0.7\times 10^{15}\,h^{-1}M_{\odot}$ .}
\label{tab:nbullet}
\end{deluxetable}
\clearpage
\begin{deluxetable}{ccc}
\tablewidth{0pt}
\setlength{\tabcolsep}{5mm}
\tablecaption{Model and corresponding best-fit values of the mean and standard 
deviation of the Gaussian fitting function}
\tablehead{model & $(\nu,\sigma_{\nu})$\tablenotemark{a}& 
$(\nu,\sigma_{\nu})$\tablenotemark{b}} 
\startdata
$\Lambda$CDM  & $(2.99,\ 0.076)$ & $(3.03,\ 0.066)$ \\
EXP001  & $(3.00,\ 0.081)$ & $(3.04,\ 0.071)$ \\ 
EXP002 & $(3.01,\ 0.091)$ & $(3.06,\ 0.081)$ \\  
EXP003  & $(3.04,\ 0.086)$ & $(3.08,\ 0.076)$  \\
EXP008e3  & $(3.05,\ 0.091)$ & $(3.08,\ 0.091)$  \\
SUGRA003 & $(2.98,\ 0.081)$ & $(3.02,\ 0.086)$ \\  
\enddata
\tablenotetext{a}{For $M_{h}\ge 0.5\times 10^{15}\,h^{-1}M_{\odot}$.} 
\tablenotetext{b}{For $M_{h}\ge 0.7\times 10^{15}\,h^{-1}M_{\odot}$ .}
\label{tab:fit}
\end{deluxetable}

\end{document}